\documentclass[referee]{raa}           
\usepackage{graphicx,times}
\usepackage{natbib}
\usepackage{amssymb,amsmath}
\usepackage{longtable}
\bibpunct{(}{)}{;}{a}{}{,}

\usepackage[a4paper=true,dvipdfm=true,pagebackref=true]{hyperref}
\hypersetup{pdftitle = The title of my PDF, pdfauthor = My name, pdfsubject= The subject, pdfkeywords = keyword1 keyword2 keyword3} 
\hypersetup{colorlinks = true, linkcolor = green, anchorcolor = red, citecolor = blue, filecolor = red, pagecolor = red, urlcolor = red}

\begin{document}

   \title{Long-term photometric behavior of the PMS stars V977 Cep and V982 Cep in the vicinity of NGC 7129
$^*$
\footnotetext{\small $*$ Based on observations obtained at Rozhen National Astronomical Observatory, Bulgaria.}
}

 \volnopage{ {\bf 2012} Vol.\ {\bf X} No. {\bf XX}, 000--000}
   \setcounter{page}{1}

   \author{Sunay I. Ibryamov\inst{1,2}, Evgeni H. Semkov\inst{2}, Teodor R. Milanov\inst{1}, Stoyanka P. Peneva\inst{2}
   }

   \institute{ Department of Theoretical and Applied Physics, University of Shumen, BG-9712 Shumen, Bulgaria; {\it sibryamov@shu.bg}\\
        \and
             Institute of Astronomy and National Astronomical Observatory, Bulgarian Academy of Sciences, BG-1784 Sofia, Bulgaria\\
	\vs \no
   {\small Received 2016 October 14; accepted 2016 November 25}
}

\abstract{Long-term $BVRI$ photometric light curves of the pre-main sequence stars V977 Cep and V982 Cep during the period from 2000 October to 2016 August are presented. The stars are located in the vicinity of the reflection nebula NGC 7129. Our photometric data show that both stars exhibit strong photometric variability in all optical passbands, which is typical for Classical T Tauri stars. Using our observational data we analyze the reasons for the observed brightness variations. In the case of V977 Cep we identify previously unknown periodicity in its light curve. 
}

   \authorrunning{S. Ibryamov et al.}            
   \titlerunning{Long-term Photometric Behavior of V977 Cep and V982 Cep}  
   \maketitle

%

\section{Introduction}           
\label{sect:intro}

The bright reflection nebula NGC 7129 is an active star forming region, where a large number of T Tauri stars, Herbig Ae/Be stars, Herbig-Haro objects and cometary nebulae can be found. According to \cite{Kun+etal+2009} NGC 7129 is probably associated with the Cepheus Bubble. The distance to NGC 7129 was determined as 1.15 kpc and its age is about 3 Myr (\citealt{Straizys+etal+2014}). Recent studies of young stellar objects in the vicinity of NGC 7129 were made by \cite{Kun+etal+2008}, \cite{Kun+etal+2009}, \cite{Bae+etal+2011}, \cite{Straizys+etal+2014}, \cite{Ibryamov+etal+2014}, \cite{Dahm+Hillenbrand+2015}, \cite{Movsessian+etal+2015}, etc.

T Tauri type stars are pre-main sequence (PMS) stars with relatively low mass (M $\leq$ 2M$_{\sun}$). T Tauri stars (TTSs) are still contracting towards the main sequence and convert their own gravitational potential energy to light. Unlike in main sequence stars, the temperature in the core of TTS is not sufficient to start of nuclear fusion. 

A distinctive feature of TTSs is that they are associated with dark nebulae and molecular clouds and are grouped in so called T associations. TTSs show strong irregular photometric variability and emission line spectra (\citealt{Joy+1945}). Each TTS can have different photometric variations which makes their classification purely by the shape of their light curve quite uncertain.

TTSs are separated into two sub-classes $-$ Classical T Tauri stars (CTTSs) and Weak-line T Tauri stars (WTTSs). CTTSs are surrounded by a circumstellar disk, while WTTSs show no evidence of such a disk (\citealt{Bertout+1989}). Both sub-classes of TTSs show photometric variability with different amplitudes. The variability of CTTS is often associated with variable accretion from the circumstellar disk and the existence of hot spots on the stellar surface (\citealt{Herbst+etal+2007}). The existence of cool spots or groups of spots and/or flare-like events (in the $B$- and $U$-bands) are responsible for the observed photometric variability in WTTSs.

In some PMS stars, large amplitude drops in the brightness (reaching up to 3 mag in the $V$-band) are observed. The stars with such photometric behavior are known as UXors (the name comes from their prototype UX Orionis). The observed drops in the star's brightness can last for several days and in some cases a few weeks. These drops are probably caused by obscuration of the young star from circumstellar clouds or dust in the line of sight to the star (see \citealt{Herbst+etal+2007} and \citealt{Dullemond+etal+2003}). In the very deep minima, the color indices of UXors often becomes bluer; this is the so called 'blueing effect' (see \citealt{Bibo+The+1990}).

The stars from our study, V977 Cep and V982 Cep, are located in the vicinity of NGC 7129. Long-term multicolor observations of PMS stars are important for their exact classification. Photometric information, especially concerning long-term behavior of the stars from our study, is missing in literature. 

Variability in V977 Cep (2MASS J21402965+6626442) was mentioned in \cite{Popov+etal+2011}, where photometric data of the star in the $R$-band (for the period 2010 February 03$-$March 02) and the $I$-band (for the period 2009 October 22$-$2010 February 21) are given. The authors provide a finding map of the star.

The star V982 Cep (2MASS J21413315+6622204) was included in the list of young stellar objects candidates with H$\alpha$ in emission in the study of \cite{Kun+1998}. \cite{Kun+etal+2009} measured $B$=17.22, $V$=15.67, $R$=14.61 and $I$=13.57 magnitudes of V982 Cep. The authors defined the spectral type of the star as K4 and determined its mass as 1.60 M$_{\sun}$, its effective temperature as 4590 K, and its age as 0.5 Myr. The spectrum of V982 Cep includes H$\alpha$, [OI] 6300, OI 7773, 8446 and CaII triplet emission lines, and the star was classified as a CTTS (\citealt{Kun+etal+2009}). \cite{Popov+etal+2011} presented photometric data in the $V$-band (for the period 2009 October 22$-$2010 February 21), the $R$-band (for the period 2010 February 03$-$2010 March 02) and the $I$-band (for the period 2009 October 22$-$2010 February 21) of V982 Cep and provided a finding map.

Section 2 in the present paper gives information about our photometric observations, telescopes and cameras used and data reduction. Section 3 describes the obtained results and their interpretation.


\section{Observations and Data reduction}
\label{sect:Obs}

The $BVRI$ photometric observations of the stars from our study were performed during the period from 2000 October 30 to 2016 August 06. The observations were carried out with the 50/70-cm Schmidt and the 60-cm Cassegrain telescopes administered by Rozhen National Astronomical Observatory in Bulgaria. 

\newpage
The observations were performed with four different types of CCD cameras: SBIG ST-8, SBIG STL-11000M and FLI PL16803 on the 50/70-cm Schmidt telescope, and FLI PL09000 on the 60-cm Cassegrain telescope. The technical parameters and specifications for the telescopes and CCD cameras used are given in \cite{Ibryamov+etal+2014}. 

All frames were taken through a standard Johnson-Cousins set of filters. All obtained frames are dark frame subtracted and flat field corrected. The photometric data were reduced using subroutine \textsc{daophot} in the \textsc{idl} software package. All data were analyzed using the same aperture, which was chosen to have a 6$\arcsec$ radius and background annulus from 10$\arcsec$ to 15$\arcsec$. As a reference sequence we used the $BVRI$ comparisons reported in \cite{Semkov+2002, Semkov+2003}. The average value of the errors in the reported magnitudes are 0.01-0.02 mag for the $I$- and $R$-band data, and 0.02-0.03 mag for the $V$- and $B$-band data.

\section{Results and Discussion}

\begin{figure}
   \centering
   \includegraphics[width=12.0cm, angle=0]{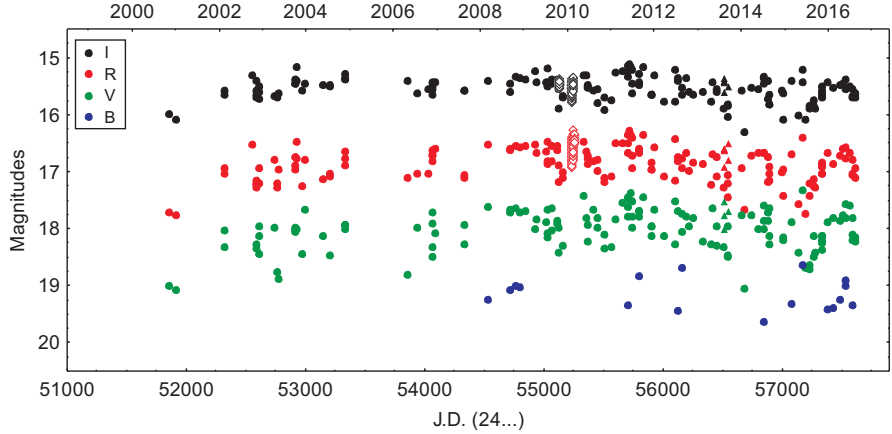}
   \caption{$IRVB$ light curves of V977 Cep during the period 2000 October$-$2016 August.} 
   \label{Fig1}
   \end{figure}

The results from our long-term $BVRI$ CCD observations of V977 Cep are summarized in Table~\ref{Tab1}. The table contains date (YYYYMMDD format) and Julian date (J.D.) of the observations, $IRVB$ magnitudes of the star, telescope and CCD camera used. The available $BVRI$ photometric data of the star V977 Cep are plotted in Figure~\ref{Fig1}. In the figure, circles represent CCD photometric data taken with the 50/70-cm Schmidt telescope, triangles mark the photometric data obtained with the 60-cm Cassegrain telescope and empty diamonds signify the photometric data from \cite{Popov+etal+2011}. 

The data reported in the paper indicate that during our study the brightness of V977 Cep varies around some intermediate level. The star's brightness during the whole observational period varies in the range 15.10--16.30 mag for the $I$-band, 16.28--17.76 mag for the $R$-band, 17.32--19.05 mag for the $V$-band and 18.63--19.44 mag for the $B$-band. For V977 Cep we only have $B$-band data when the star is near its maximum brightness due to the photometric limit of the telescopes used (19.5 mag). 

\newpage
The observed amplitudes for the period 2000--2016 are 1.20 mag for the $I$-band, 1.48 mag for the $R$-band, 1.73 mag for the $V$-band and $>$0.81 mag for the $B$-band. Variability with such amplitudes is typical of CTTS surrounded by a circumstellar disk and it is an indication for the presence of variable accretion from the circumstellar disk onto the stellar surface.

\begin{figure}
\begin{center}
\includegraphics[width=3.5cm]{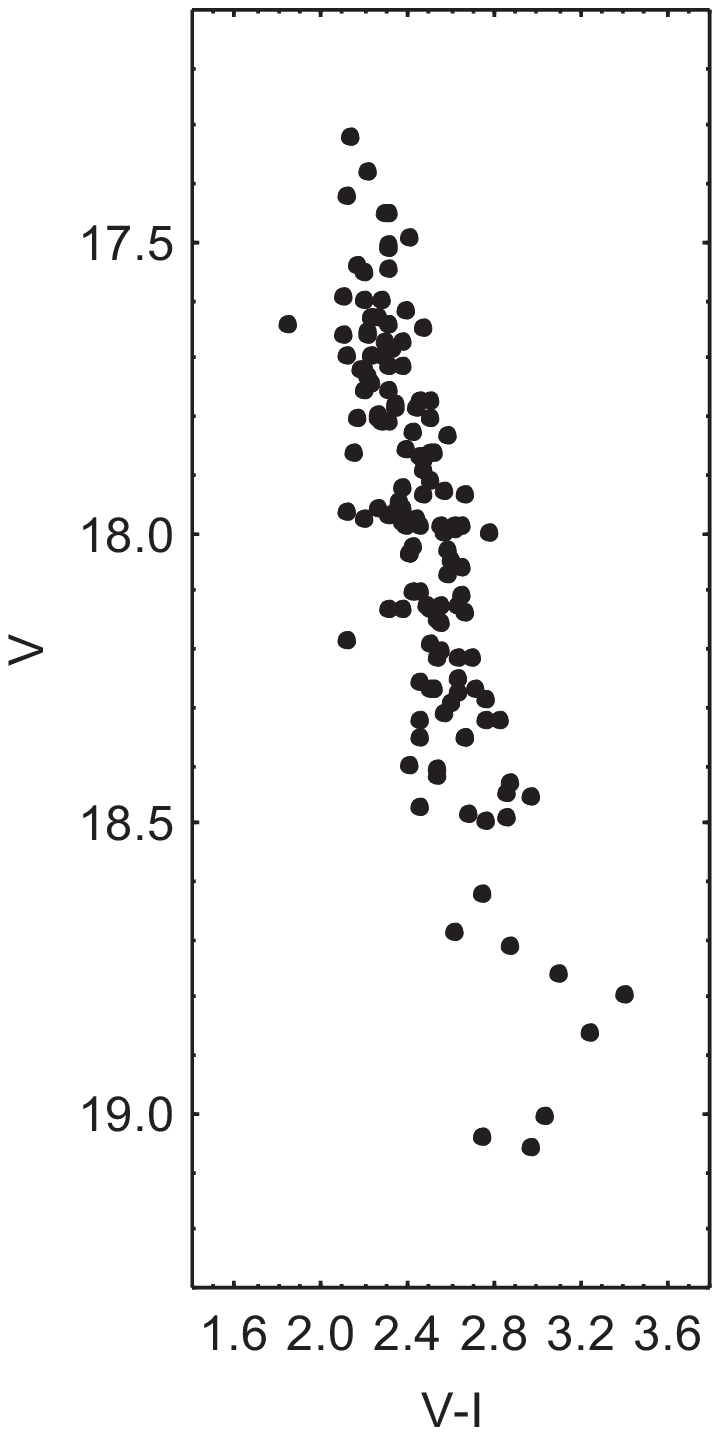}
\includegraphics[width=3.5cm]{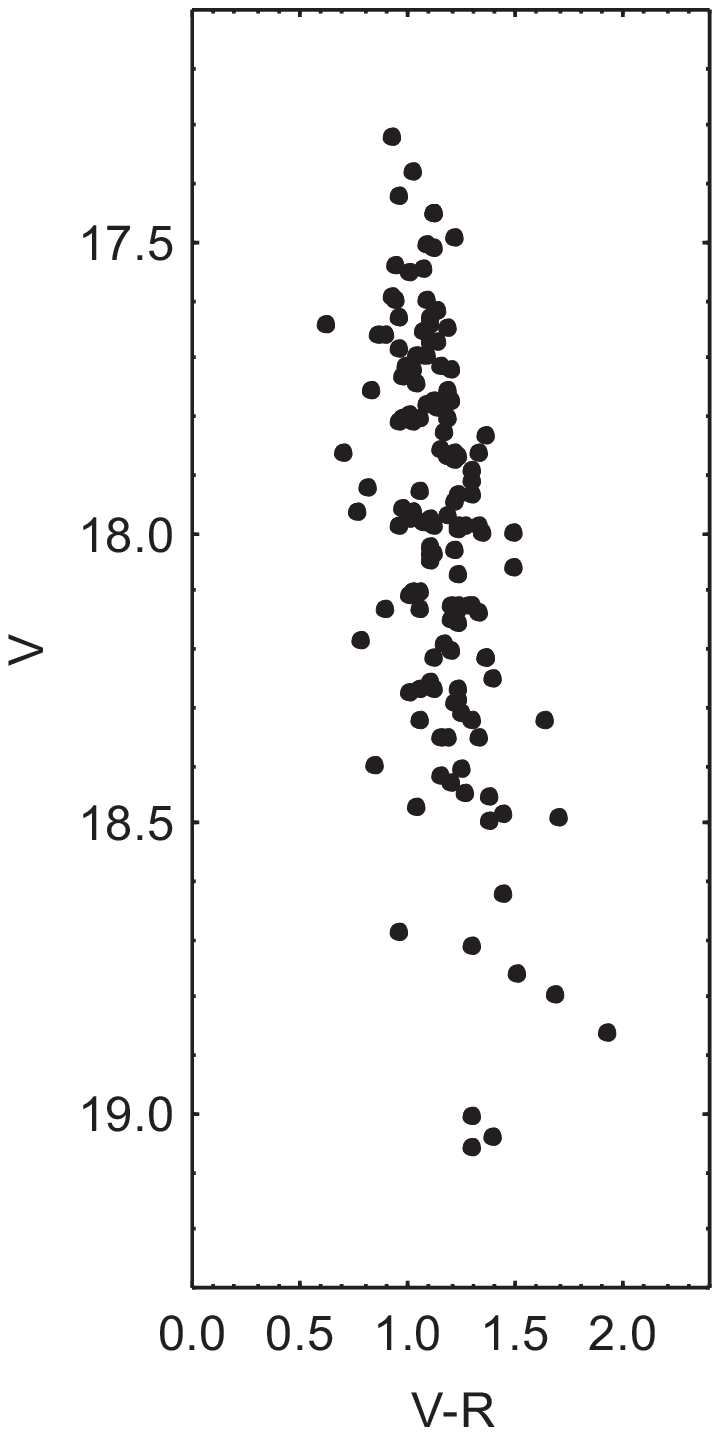}
\caption{Color indices $V-I$ and $V-R$ vs. the stellar $V$ magnitude of V977 Cep during the period 2000 October$-$2016 August.}\label{Fig2}
\end{center}
\end{figure}

The measured color indices $V-I$ and $V-R$ vs. the stellar $V$ magnitude of V977 Cep during the period of our observations are plotted on Figure~\ref{Fig2}. It can be seen that the star becomes redder as it fades, and the blueing effect is not observed. Such color variations are typical for both CTTSs and WTTSs, whose variability is produced by the rotational modulation of spots on the stellar surface.

\begin{figure}
   \centering
   \includegraphics[width=12.0cm, angle=0]{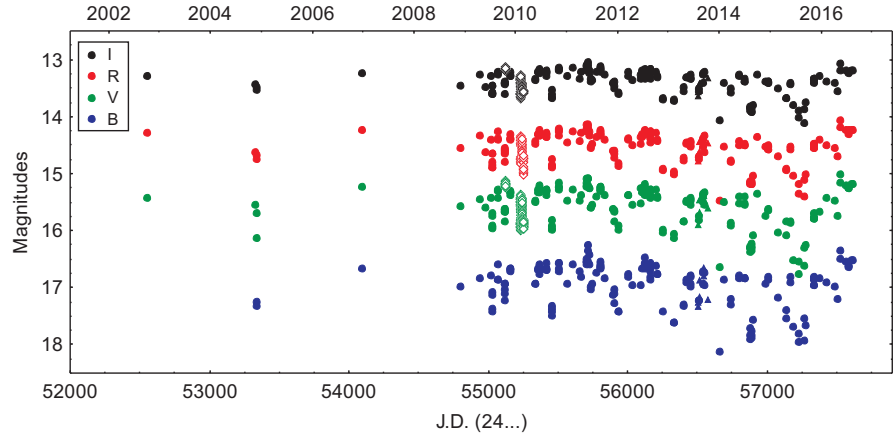}
   \caption{$IRVB$ light curves of V982 Cep during the period 2002 October--2016 August.} 
   \label{Fig3}
   \end{figure}

V982 Cep is located at about 7$\arcmin$ 45$\arcsec$ from V977 Cep and at about 18$\arcmin$ from the center of NGC 7129. Figure~\ref{Fig3} shows $BVRI$ light curves of V982 Cep. The symbols used are the same as in Figure~\ref{Fig1}. The results of our long-term CCD observations are summarized in Table~\ref{Tab2}. The columns have the same contents as in Table~\ref{Tab1}. During our observations, the star's brightness varies in the range 13.04--14.10 mag for the $I$-band, 15.46--14.05 mag for the $R$-band, 14.99--16.75 mag for the $V$-band and 16.23--18.10 mag for the $B$-band. The observed amplitudes are 1.06 mag for the $I$-band, 1.41 mag for the $R$-band, 1.76 mag for the $V$-band and 1.87 mag for the $B$-band.

From Figure~\ref{Fig3} we can see that for most of our CCD observations V982 Cep is at high brightness. The star shows irregular fading events in all bands with different amplitudes and durations. The fading events with larger amplitudes in the brightness of the star were observed in 2009 July, 2010 September, 2011 December, 2013 February, 2013 December and 2014 August. The deepest and most prolonged drop in the brightness of V982 Cep was observed in 2015 August ($\Delta I$=0.90 mag, $\Delta R$=1.08 mag, $\Delta V$=1.42 mag, $\Delta B$=1.28 mag). One could suggest that fading events happen frequently during periods with missing data.

\begin{figure}
\begin{center}
\includegraphics[width=3.5cm]{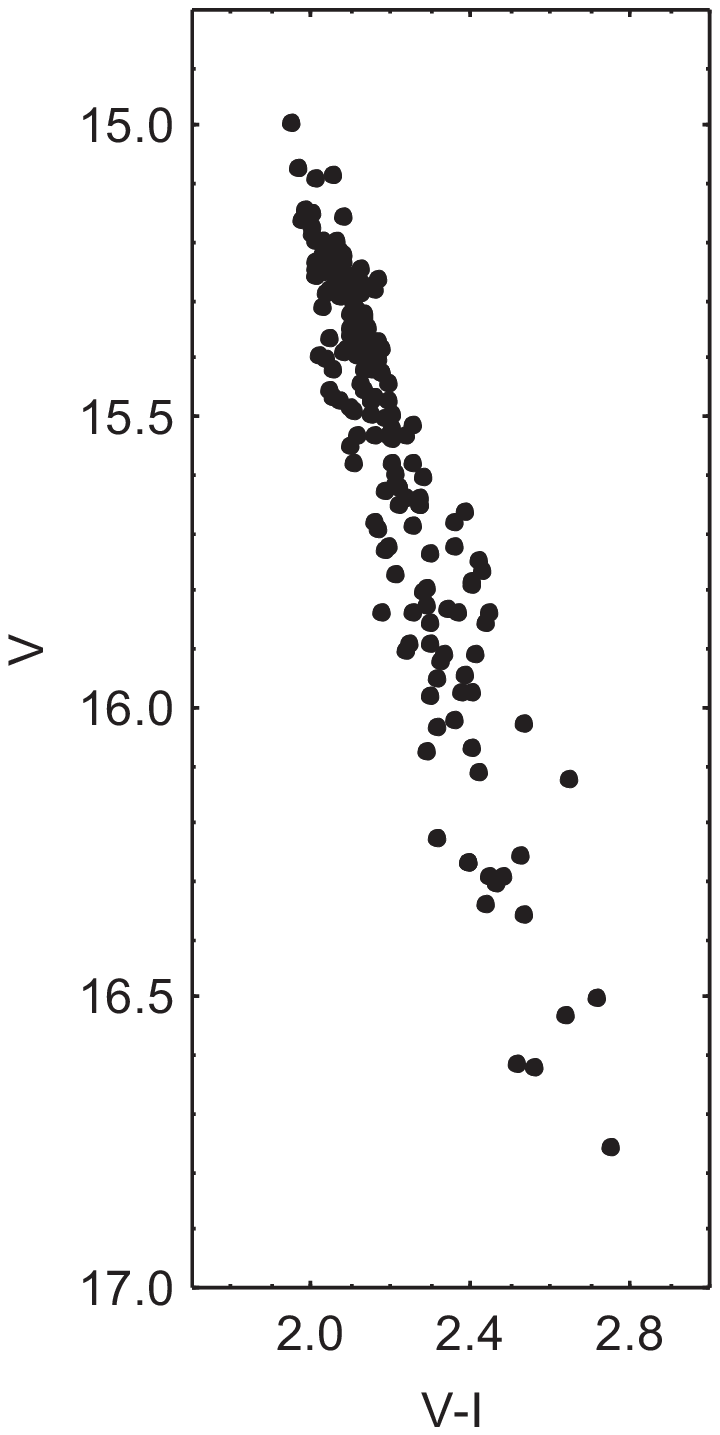}
\includegraphics[width=3.5cm]{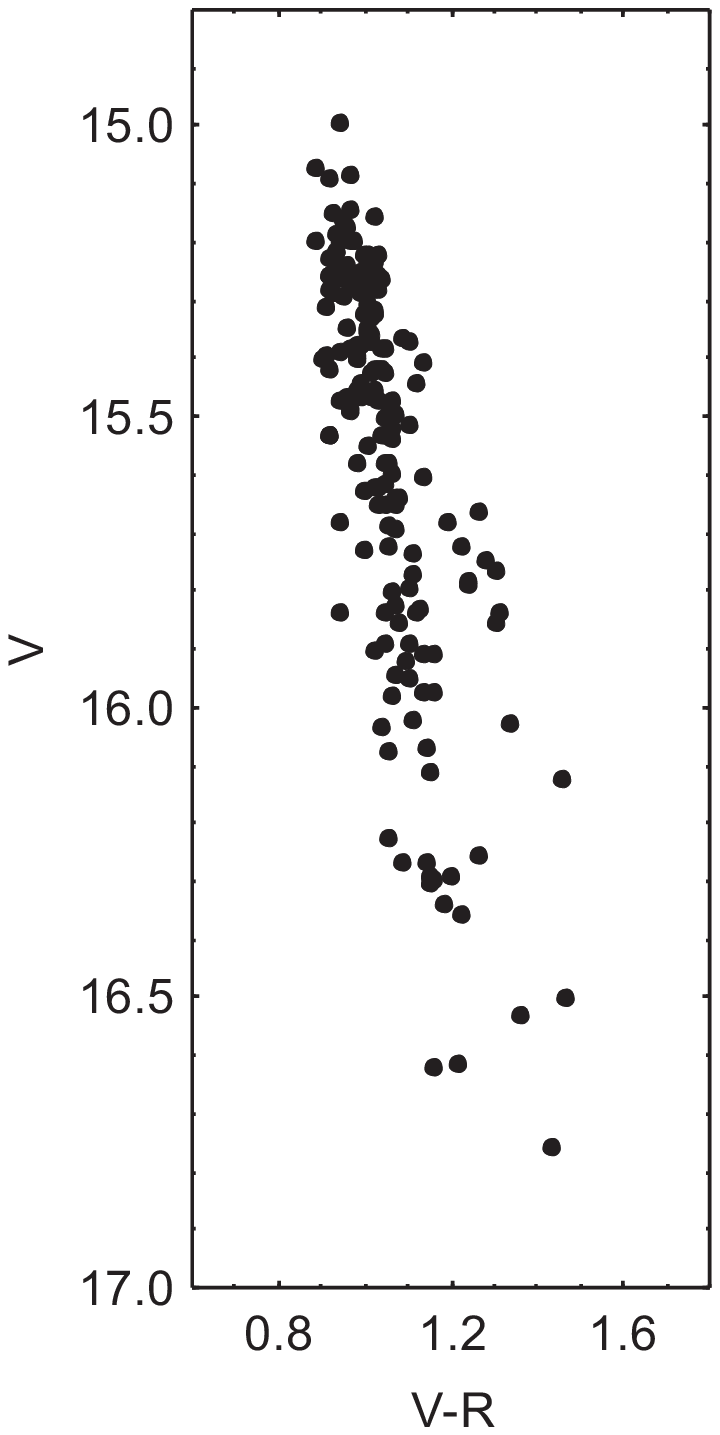}
\includegraphics[width=3.5cm]{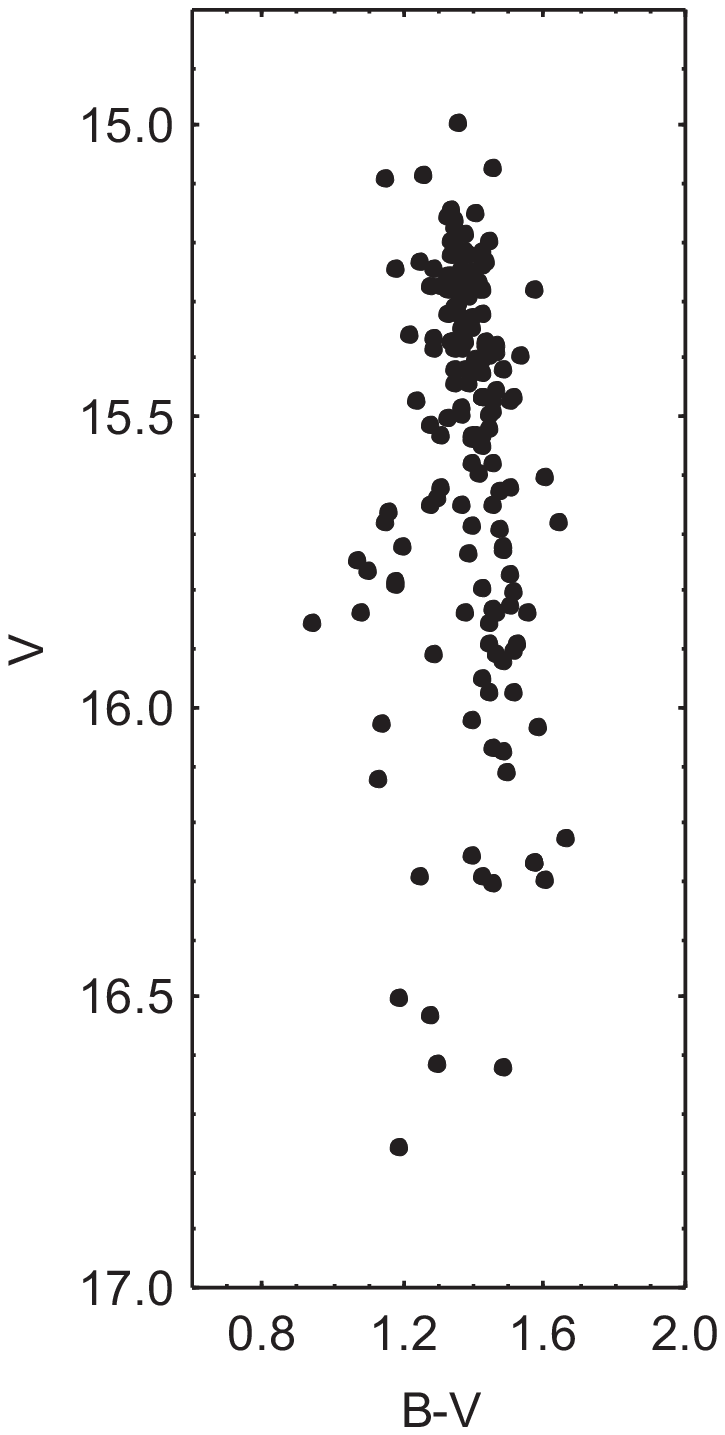}
\caption{Color indices $V-I$, $V-R$ and $B-V$ vs. the stellar $V$ magnitude of V982 Cep during the period 2002 October$-$2016 August.}\label{Fig4}
\end{center}
\end{figure}

\newpage
The measured color indices $V-I$, $V-R$ and $B-V$ vs. the stellar $V$ magnitude of V982 Cep during the period of our observations are plotted in Figure~\ref{Fig4}. The figure shows evidence of the blueing effect, which is most obvious for the $B-V$ index vs the stellar $V$ magnitude. The amplitudes of the observed drops in the brightness of the star and the existence of the blueing effect are indications of UXor type variability. It is likely that the observed fading events in the light curve of V982 Cep are due to obscuration from circumstellar clouds of protostellar material and/or the existence of planetesimals around the star.

We used the 2MASS $JHK_{s}$ magnitudes of V977 Cep and V982 Cep to construct a two-color diagram to check whether the stars have infrared excess, which is an indication for the presence of a circumstellar disk.

\begin{figure}
   \centering
   \includegraphics[width=8.0cm, angle=0]{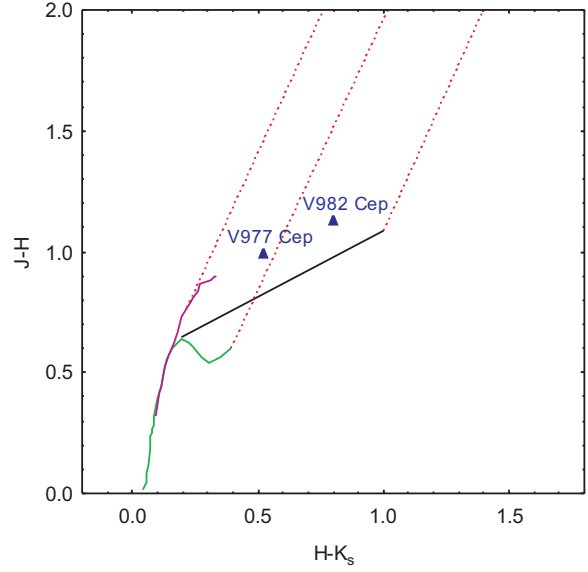}
   \caption{The $J-H$ vs. $H-K_{s}$ diagram for V977 Cep and V982 Cep detected in $J$-, $H$- and $K_{s}$-bands in the 2MASS catalog.} 
   \label{Fig5}
   \end{figure}

\newpage
Figure~\ref{Fig5} shows the location of main sequence (green line) and giant stars (purple line) from \cite{Bessell+Brett+1988}, and the location of CTTSs (black line) from \cite{Meyer+etal+1997}. A correction to the 2MASS photometric system was performed following the procedure in \cite{Carpenter+2001}. The three red parallel dotted lines show the direction of the interstellar reddening vectors determined for the NGC 7129 region by \cite{Straizys+etal+2014}.

From Figure~\ref{Fig5} we can see that both V977 Cep and V982 Cep lie about 0.2 mag above the intrinsic CTTS line. Therefore the stars from our study have clear infrared excess, indicating the presence of disks around them. All existing photometric and spectral data for V977 Cep and V982 Cep suggest that they can be classified as CTTSs. Additionally, V982 Cep exhibits UXor type variability, which is known to be inherent in TTSs.

We used the software packages \textsc{persea} (\citealt{Schwarzenberg-Czerny+1996}) and \textsc{period04} (\citealt{Lenz+Breger+2005}) to search for periodicity in the light curves of the stars from our study. We did not identify any periodicity in the variations of V982 Cep, however, our time-series analysis of V977 Cep covering the period 2010--2016 showed a 8.149 $\pm$ 0.038 d period and led to the ephemeris

\begin{equation}\label{eq1}
JD (max) = 2456007.444384 + 8.149221 * E.
\end{equation}

\begin{figure}
   \centering
   \includegraphics[width=8.0cm, angle=0]{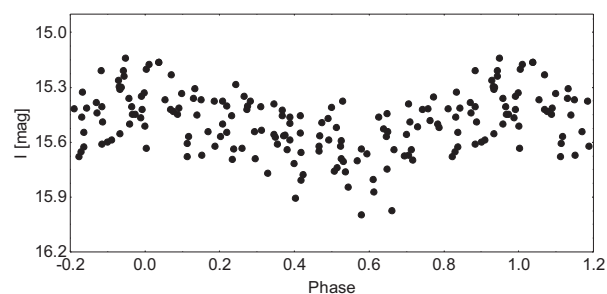}
   \caption{$I$-band folded light curve of V977 Cep.} 
   \label{Fig6}
   \end{figure}

False Alarm Probability (FAP) estimation was done by randomly deleting about 15$\%$ of the data about 10 times and then redetermining the period. The period and starting age determinations remained stable even when a subsample of about 20$\%$ of the data were removed. Figure~\ref{Fig6} shows the $I$-band folded light curve of V977 Cep according to the ephemeris (1). The data obtained in $RVB$-bands show the same shape of the folded light curves.

\newpage
The discovered periodicity in the light curve of V977 Cep is stable during a time interval of several years. It is a typical rotational period for CTTSs (see \citealt{Bouvier+etal+1995}). The periodicity could be caused by rotational modulation of spots on the stellar surface. Unlike WTTSs, which have rotational periods in the range 2-5 d, CTTSs have rotational periods in the range 6-9 d. According to \cite{Petrov+2003}, it is possible that the existence of an accreting disk and stellar wind in CTTS somehow slows down its rotation.

\section{Conclusion}

The long-term multicolor light curves of the PMS stars V977 Cep and V982 Cep during the period 2000 to 2016 are presented and discussed. Both stars show photometric characteristics of CTTSs. The observed amplitudes in their brightness variations, the shapes of their light curves and the found period of 8.149 d for V977 Cep confirmed that suspicion. It is highly likely that V982 Cep has a UXor type variability. We found evidence for the blueing effect in its color-stellar magnitude diagram. Further photometric and spectral observations of the stars from our study will be of great importance for determining their exact classification.

\normalem
\begin{acknowledgements}
This work was supported partly by funds of the project 'Multicolor photometric study of Pre-main sequence stars from selected star-forming regions' financed by Fund for Scientific Research of the Bulgarian Ministry of Education and Science. This research has made use of the NASA's Astrophysics Data System Abstract Service, the SIMBAD database and the VizieR catalogue access tool, operated at CDS, Strasbourg, France. This publication makes use of data products from the Two Micron All Sky Survey, which is a joint project of the University of Massachusetts and the Infrared Processing and Analysis Center/California Institute of Technology, funded by the National Aeronautics and Space Administration and the National Science Foundation (\citealt{Skrutskie+etal+2006}). This research was supported partly by funds of the project RD-08-81 at the University of Shumen.
\end{acknowledgements}
  	
\newpage
\bibliographystyle{raa}
\bibliography{bibtex}


\newpage
{\small
\begin{longtable}{cccccccc}
\caption{Photometric CCD Observations and Data of V977 Cep during the Period 2000 October$-$2016 August}\\
\hline\hline
\noalign{\smallskip}  
Date \hspace{1.5cm} &	J.D. (24...) \hspace{2mm}	&	$I$ [mag]	\hspace{4mm} & $R$ [mag] \hspace{4mm} & $V$ [mag] \hspace{4mm} & $B$ [mag] \hspace{4mm} & Telescope \hspace{1mm} & CCD camera\\
\noalign{\smallskip}  
\hline
\endfirsthead
\caption{continued.}\\
\hline\hline
\noalign{\smallskip}  
Date \hspace{1.5cm} &	J.D. (24...) \hspace{2mm}	&	$I$ [mag]	\hspace{4mm} & $R$ [mag] \hspace{4mm} & $V$ [mag] \hspace{4mm} & $B$ [mag] \hspace{4mm} & Telescope \hspace{1mm} & CCD camera\\
\noalign{\smallskip}  
\hline
\noalign{\smallskip}  
\endhead
\hline
\label{Tab1}
\endfoot
\noalign{\smallskip}
20001030	&	51848.412	&	15.97	&	17.71	&	19.00	&	-	&	Schmidt	&	ST-8	\\
20001224	&	51903.282	&	16.09	&	17.76	&	19.05	&	-	&	Schmidt	&	ST-8	\\
20020206	&	52312.248	&	15.63	&	16.94	&	18.03	&	-	&	Schmidt	&	ST-8	\\
20020207	&	52313.233	&	15.56	&	17.03	&	18.32	&	-	&	Schmidt	&	ST-8	\\
20021003	&	52551.459	&	15.31	&	16.51	&	-	&	-	&	Schmidt	&	ST-8	\\
20021029	&	52577.405	&	15.57	&	17.14	&	-	&	-	&	Schmidt	&	ST-8	\\
20021030	&	52578.358	&	15.69	&	17.21	&	18.35	&	-	&	Schmidt	&	ST-8	\\
20021031	&	52579.262	&	15.39	&	17.19	&	-	&	-	&	Schmidt	&	ST-8	\\
20021101	&	52580.246	&	15.65	&	17.27	&	18.27	&	-	&	Schmidt	&	ST-8	\\
20021126	&	52605.224	&	15.70	&	-	&	17.95	&	-	&	Schmidt	&	ST-8	\\
20021128	&	52607.281	&	15.50	&	16.93	&	18.12	&	-	&	Schmidt	&	ST-8	\\
20021129	&	52608.233	&	15.60	&	17.19	&	18.44	&	-	&	Schmidt	&	ST-8	\\
20030403	&	52732.524	&	15.67	&	16.78	&	17.97	&	-	&	Schmidt	&	ST-8	\\
20030501	&	52761.475	&	15.67	&	17.26	&	18.76	&	-	&	Schmidt	&	ST-8	\\
20030502	&	52762.437	&	15.68	&	17.20	&	-	&	-	&	Schmidt	&	ST-8	\\
20030505	&	52765.408	&	15.62	&	16.95	&	18.86	&	-	&	Schmidt	&	ST-8	\\
20030927	&	52910.301	&	15.45	&	16.95	&	18.04	&	-	&	Schmidt	&	ST-8	\\
20030928	&	52911.283	&	15.45	&	16.72	&	17.98	&	-	&	Schmidt	&	ST-8	\\
20030929	&	52912.273	&	15.37	&	16.88	&	17.98	&	-	&	Schmidt	&	ST-8	\\
20031002	&	52915.341	&	15.15	&	16.46	&	-	&	-	&	Schmidt	&	ST-8	\\
20031003	&	52916.318	&	15.39	&	16.77	&	17.99	&	-	&	Schmidt	&	ST-8	\\
20031125	&	52969.263	&	15.57	&	17.24	&	18.43	&	-	&	Schmidt	&	ST-8	\\
20031219	&	52993.201	&	15.45	&	16.77	&	17.66	&	-	&	Schmidt	&	ST-8	\\
20040513	&	53138.651	&	15.46	&	17.11	&	18.11	&	-	&	Schmidt	&	ST-8	\\
20040715	&	53201.571	&	15.49	&	17.08	&	18.45	&	-	&	Schmidt	&	ST-8	\\
20040716	&	53202.574	&	15.46	&	17.02	&	-	&	-	&	Schmidt	&	ST-8	\\
20041117	&	53327.401	&	15.37	&	16.88	&	17.92	&	-	&	Schmidt	&	ST-8	\\
20041118	&	53328.350	&	15.35	&	16.75	&	17.99	&	-	&	Schmidt	&	ST-8	\\
20041120	&	53330.375	&	15.27	&	16.65	&	17.93	&	-	&	Schmidt	&	ST-8	\\
20060424	&	53849.515	&	15.40	&	17.11	&	18.79	&	-	&	Schmidt	&	ST-8	\\
20060719	&	53936.431	&	15.61	&	17.04	&	17.98	&	-	&	Schmidt	&	ST-8	\\
20061020	&	54029.411	&	15.54	&	17.02	&		&	-	&	Schmidt	&	ST-8	\\
20061117	&	54057.271	&	15.64	&	16.80	&	18.49	&	-	&	Schmidt	&	ST-8	\\
20061118	&	54058.279	&	15.54	&	16.70	&	17.72	&	-	&	Schmidt	&	ST-8	\\
20061119	&	54059.274	&	15.50	&	16.69	&	18.32	&	-	&	Schmidt	&	ST-8	\\
20061120	&	54060.227	&	15.43	&	16.60	&	17.89	&	-	&	Schmidt	&	ST-8	\\
20061216	&	54086.277	&	15.42	&	16.58	&	18.06	&	-	&	Schmidt	&	ST-8	\\
20070818	&	54331.353	&	15.56	&	17.11	&	17.92	&	-	&	Schmidt	&	ST-8	\\
20070819	&	54332.343	&	15.56	&	17.04	&	18.27	&	-	&	Schmidt	&	ST-8	\\
20080229	&	54526.644	&	15.40	&	16.51	&	17.60	&	19.24	&	Schmidt	&	STL-11	\\
20080827	&	54706.388	&	15.45	&	16.59	&	17.65	&	-	&	Schmidt	&	STL-11	\\
20080828	&	54707.498	&	15.58	&	16.61	&	17.69	&	19.06	&	Schmidt	&	STL-11	\\
20081021	&	54761.255	&	15.33	&	16.55	&	17.64	&	18.98	&	Schmidt	&	STL-11	\\
20081120	&	54791.191	&	15.35	&	16.57	&	17.71	&	19.01	&	Schmidt	&	STL-11	\\
20090112	&	54844.234	&	15.37	&	16.54	&	17.67	&	-	&	Schmidt	&	STL-11	\\
20090326	&	54917.518	&	15.22	&	16.51	&	18.00	&	-	&	Schmidt	&	STL-11	\\
20090416	&	54938.453	&	15.41	&	16.67	&	17.82	&	-	&	Schmidt	&	STL-11	\\
20090628	&	55011.525	&	15.42	&	16.65	&	17.87	&	-	&	Schmidt	&	FLI	\\
20090714	&	55027.439	&	15.48	&	16.80	&	18.13	&	-	&	Schmidt	&	FLI	\\
20090715	&	55028.447	&	15.49	&	16.85	&	18.07	&	-	&	Schmidt	&	FLI	\\
20090716	&	55029.453	&	15.17	&	16.48	&	17.64	&	-	&	Schmidt	&	FLI	\\
20090821	&	55065.362	&	15.45	&	16.82	&	18.03	&	-	&	Schmidt	&	FLI	\\
20090822	&	55066.282	&	15.37	&	16.53	&	17.63	&	-	&	Schmidt	&	FLI	\\
20091006	&	55111.424	&	15.42	&	16.61	&	17.91	&	-	&	Schmidt	&	FLI	\\
20091007	&	55112.371	&	15.37	&	16.54	&	17.86	&	-	&	Schmidt	&	FLI	\\
20091008	&	55113.354	&	15.88	&	17.17	&	18.40	&	-	&	Schmidt	&	FLI	\\
20091009	&	55114.245	&	15.55	&	16.89	&	17.97	&	-	&	Schmidt	&	FLI	\\
20091120	&	55156.237	&	15.70	&	17.09	&	18.29	&	-	&	Schmidt	&	FLI	\\
20091121	&	55157.266	&	15.67	&	17.00	&	-	&	-	&	Schmidt	&	FLI	\\
20100513	&	55330.373	&	15.30	&	16.46	&	17.42	&	-	&	Schmidt	&	FLI	\\
20100608	&	55356.417	&	15.31	&	16.63	&	17.80	&	-	&	Schmidt	&	FLI	\\
20100610	&	55358.489	&	15.50	&	16.75	&	-	&	-	&	Schmidt	&	FLI	\\
20100611	&	55359.507	&	15.52	&	16.85	&	18.21	&	-	&	Schmidt	&	FLI	\\
20100612	&	55360.441	&	15.46	&	16.70	&	17.93	&	-	&	Schmidt	&	FLI	\\
20100806	&	55415.396	&	15.56	&	16.80	&	17.65	&	-	&	Schmidt	&	FLI	\\
20100807	&	55447.516	&	15.53	&	16.79	&	17.80	&	-	&	Schmidt	&	FLI	\\
20100908	&	55448.434	&	15.78	&	16.97	&	17.97	&	-	&	Schmidt	&	FLI	\\
20101104	&	55505.299	&	15.90	&	17.17	&	18.35	&	-	&	Schmidt	&	FLI	\\
20101105	&	55506.310	&	15.69	&	17.09	&	18.10	&	-	&	Schmidt	&	FLI	\\
20110101	&	55563.257	&	15.74	&	17.07	&	18.31	&	-	&	Schmidt	&	FLI	\\
20110206	&	55599.228	&	15.26	&	16.48	&	17.83	&	-	&	Schmidt	&	FLI	\\
20110404	&	55656.398	&	15.24	&	16.48	&	17.54	&	-	&	Schmidt	&	FLI	\\
20110522	&	55704.393	&	15.33	&	16.66	&	17.77	&	19.34	&	Schmidt	&	FLI	\\
20110523	&	55705.334	&	15.14	&	16.33	&	17.45	&	-	&	Schmidt	&	FLI	\\
20110524	&	55706.323	&	15.23	&	16.49	&	17.62	&	-	&	Schmidt	&	FLI	\\
20110525	&	55707.335	&	15.10	&	16.28	&	17.49	&	-	&	Schmidt	&	FLI	\\
20110609	&	55722.355	&	15.16	&	16.35	&	17.37	&	-	&	Schmidt	&	FLI	\\
20110621	&	55734.469	&	15.62	&	16.92	&	17.98	&	-	&	Schmidt	&	FLI	\\
20110622	&	55735.475	&	15.44	&	16.70	&	17.78	&	-	&	Schmidt	&	FLI	\\
20110624	&	55737.404	&	15.21	&	16.39	&	17.50	&	-	&	Schmidt	&	FLI	\\
20110823	&	55797.367	&	15.28	&	16.58	&	17.77	&	-	&	Schmidt	&	FLI	\\
20110824	&	55798.391	&	15.42	&	16.62	&	17.69	&	-	&	Schmidt	&	FLI	\\
20110825	&	55799.393	&	15.47	&	16.66	&	17.69	&	18.82	&	Schmidt	&	FLI	\\
20110923	&	55828.303	&	15.16	&	16.34	&	17.45	&	-	&	Schmidt	&	FLI	\\
20111127	&	55893.206	&	15.63	&	16.94	&	17.96	&	-	&	Schmidt	&	FLI	\\
20111129	&	55895.297	&	15.63	&	16.94	&	17.96	&	-	&	Schmidt	&	FLI	\\
20111130	&	55896.278	&	15.64	&	16.83	&	18.12	&	-	&	Schmidt	&	FLI	\\
20111229	&	55925.222	&	15.30	&	16.56	&	17.67	&	-	&	Schmidt	&	FLI	\\
20120316	&	56003.489	&	15.76	&	17.08	&	18.13	&	-	&	Schmidt	&	FLI	\\
20120612	&	56091.442	&	15.77	&	17.15	&	18.26	&	-	&	Schmidt	&	FLI	\\
20120617	&	56096.448	&	15.21	&	16.42	&	17.50	&	-	&	Schmidt	&	FLI	\\
20120711	&	56120.421	&	15.40	&	16.81	&	-	&	-	&	Schmidt	&	FLI	\\
20120713	&	56122.422	&	15.36	&	16.73	&	17.68	&	19.44	&	Schmidt	&	FLI	\\
20120819	&	56159.394	&	15.64	&	16.92	&	18.03	&	-	&	Schmidt	&	FLI	\\
20120820	&	56160.375	&	15.52	&	16.76	&	17.73	&	18.68	&	Schmidt	&	FLI	\\
20120923	&	56194.365	&	15.36	&	16.65	&	17.78	&	-	&	Schmidt	&	FLI	\\
20121009	&	56210.265	&	15.59	&	16.98	&	17.95	&	-	&	Schmidt	&	FLI	\\
20121118	&	56250.355	&	15.65	&	16.83	&	17.80	&	-	&	Schmidt	&	FLI	\\
20130204	&	56328.260	&	15.58	&	16.85	&	18.21	&	-	&	Schmidt	&	FLI	\\
20130411	&	56394.411	&	15.76	&	17.21	&	18.26	&	-	&	Schmidt	&	FLI	\\
20130502	&	56415.467	&	15.41	&	16.68	&	17.63	&	-	&	Schmidt	&	FLI	\\
20130530	&	56443.475	&	15.59	&	16.74	&	17.94	&	-	&	Schmidt	&	FLI	\\
20130531	&	56444.481	&	15.54	&	17.07	&	18.28	&	-	&	Schmidt	&	FLI	\\
20130804	&	56509.370	&	15.87	&	17.27	&	18.32	&	-	&	Schmidt	&	FLI	\\
20130805	&	56510.432	&	15.42	&	16.64	&	-	&	-	&	60-cm	&	FLI	\\
20130806	&	56511.474	&	15.38	&	16.61	&	17.54	&	-	&	60-cm	&	FLI	\\
20130807	&	56512.461	&	15.42	&	16.61	&	-	&	-	&	60-cm	&	FLI	\\
20130808	&	56513.450	&	15.61	&	16.93	&	18.02	&	-	&	60-cm	&	FLI	\\
20130809	&	56514.408	&	15.45	&	16.63	&	17.78	&	-	&	60-cm	&	FLI	\\
20130904	&	56540.378	&	15.80	&	17.05	&	18.48	&	-	&	Schmidt	&	FLI	\\
20130905	&	56541.402	&	15.84	&	17.19	&	17.96	&	-	&	Schmidt	&	FLI	\\
20130906	&	56542.442	&	16.03	&	17.44	&	18.47	&	-	&	Schmidt	&	FLI	\\
20130911	&	56547.351	&	-	&	16.80	&	-	&	-	&	60-cm	&	FLI	\\
20130914	&	56550.330	&	15.52	&	16.52	&	17.71	&	-	&	60-cm	&	FLI	\\
20131229	&	56656.327	&	15.57	&	16.94	&	17.75	&	-	&	Schmidt	&	FLI	\\
20140123	&	56681.265	&	16.30	&	17.66	&	19.03	&	-	&	Schmidt	&	FLI	\\
20140321	&	56738.504	&	15.51	&	16.71	&	17.74	&	-	&	Schmidt	&	FLI	\\
20140521	&	56799.542	&	15.54	&	16.67	&	17.98	&	-	&	Schmidt	&	FLI	\\
20140628	&	56837.445	&	15.44	&	16.66	&	18.00	&	-	&	Schmidt	&	FLI	\\
20140629	&	56838.423	&	15.35	&	16.66	&	17.86	&	-	&	Schmidt	&	FLI	\\
20140629	&	56838.478	&	15.33	&	16.65	&	17.59	&	-	&	Schmidt	&	FLI	\\
20140803	&	56873.331	&	15.40	&	16.73	&	17.71	&	-	&	Schmidt	&	FLI	\\
20140804	&	56874.353	&	15.72	&	17.17	&	17.86	&	-	&	Schmidt	&	FLI	\\
20140818	&	56888.383	&	15.81	&	17.15	&	18.25	&	-	&	Schmidt	&	FLI	\\
20140819	&	56889.320	&	15.64	&	16.90	&	18.13	&	-	&	Schmidt	&	FLI	\\
20140822	&	56892.358	&	15.80	&	17.03	&	17.64	&	-	&	Schmidt	&	FLI	\\
20141126	&	56988.238	&	15.56	&	17.01	&	-	&	-	&	Schmidt	&	FLI	\\
20141213	&	57005.283	&	15.62	&	16.96	&	18.14	&	-	&	Schmidt	&	FLI	\\
20141214	&	57006.342	&	16.07	&	17.40	&	18.18	&	-	&	Schmidt	&	FLI	\\
20150220	&	57074.535	&	15.40	&	16.69	&	17.87	&	19.31	&	Schmidt	&	FLI	\\
20150423	&	57136.579	&	16.00	&	17.56	&	18.40	&	-	&	Schmidt	&	FLI	\\
20150519	&	57162.488	&	15.41	&	-	&	-	&	-	&	Schmidt	&	FLI	\\
20150521	&	57164.501	&	15.20	&	16.40	&	17.32	&	18.62	&	Schmidt	&	FLI	\\
20150612	&	57186.506	&	16.07	&	17.73	&	18.68	&	-	&	Schmidt	&	FLI	\\
20150716	&	57220.424	&	15.84	&	17.42	&	18.71	&	-	&	Schmidt	&	FLI	\\
20150717	&	57221.478	&	15.88	&	17.19	&	18.62	&	-	&	Schmidt	&	FLI	\\
20150824	&	57259.394	&	15.74	&	17.13	&	18.49	&	-	&	Schmidt	&	FLI	\\
20150825	&	57260.384	&	15.83	&	17.23	&	18.13	&	-	&	Schmidt	&	FLI	\\
20150903	&	57269.381	&	15.89	&	17.27	&	18.41	&	-	&	Schmidt	&	FLI	\\
20151103	&	57330.285	&	15.57	&	16.84	&	18.12	&	-	&	Schmidt	&	FLI	\\
20151104	&	57331.297	&	15.49	&	16.89	&	18.12	&	-	&	Schmidt	&	FLI	\\
20151105	&	57332.286	&	15.66	&	17.01	&	18.20	&	-	&	Schmidt	&	FLI	\\
20151106	&	57333.287	&	15.70	&	17.02	&	18.35	&	-	&	Schmidt	&	FLI	\\
20151107	&	57334.267	&	15.62	&	16.86	&	18.25	&	-	&	Schmidt	&	FLI	\\
20151215	&	57372.270	&	15.41	&	16.67	&	17.87	&	19.41	&	Schmidt	&	FLI	\\
20160206	&	57425.240	&	15.51	&	16.85	&	17.81	&	19.39	&	Schmidt	&	FLI	\\
20160406	&	57485.448	&	15.46	&	16.72	&	17.85	&	19.24	&	Schmidt	&	FLI	\\
20160427	&	57506.511	&	15.44	&	16.58	&	17.75	&	-	&	Schmidt	&	FLI	\\
20160513	&	57522.485	&	15.55	&	16.75	&	17.80	&	18.89	&	Schmidt	&	FLI	\\
20160514	&	57523.471	&	15.36	&	16.55	&	17.55	&	19.00	&	Schmidt	&	FLI	\\
20160625	&	57565.464	&	15.50	&	16.67	&	17.59	&	-	&	Schmidt	&	FLI	\\
20160711	&	57581.451	&	15.54	&	16.79	&	17.79	&	19.33	&	Schmidt	&	FLI	\\
20160712	&	57582.484	&	15.69	&	17.02	&	18.18	&	-	&	Schmidt	&	FLI	\\
20160713	&	57583.470	&	15.65	&	17.04	&	18.10	&	-	&	Schmidt	&	FLI	\\
20160804	&	57605.436	&	15.68	&	17.10	&	18.21	&	-	&	Schmidt	&	FLI	\\
20160806	&	57607.415	&	15.61	&	16.93	&	18.15	&	-	&	Schmidt	&	FLI	\\
\hline \hline
\end{longtable}}

{\small
\begin{longtable}{cccccccc}
\caption{Photometric CCD Observations and Data of V982 Cep during the Period 2002 October$-$2016 August}\\
\hline\hline
\noalign{\smallskip}  
Date \hspace{1.5cm} &	J.D. (24...) \hspace{2mm}	&	$I$ [mag]	\hspace{4mm} & $R$ [mag] \hspace{4mm} & $V$ [mag] \hspace{4mm} & $B$ [mag] \hspace{4mm} & Telescope \hspace{1mm} & CCD camera\\
\noalign{\smallskip}  
\hline
\endfirsthead
\caption{continued.}\\
\hline\hline
\noalign{\smallskip}  
Date \hspace{1.5cm} &	J.D. (24...) \hspace{2mm}	&	$I$ [mag]	\hspace{4mm} & $R$ [mag] \hspace{4mm} & $V$ [mag] \hspace{4mm} & $B$ [mag] \hspace{4mm} & Telescope \hspace{1mm} & CCD camera\\
\noalign{\smallskip}  
\hline
\noalign{\smallskip}  
\endhead
\hline
\label{Tab2}
\endfoot
\noalign{\smallskip}
20021003	&	52551.459	&	13.27	&	14.27	&	15.41	&	-	&	Schmidt	&	ST8	\\
20041117	&	53327.401	&	13.42	&	14.61	&	15.53	&	-	&	Schmidt	&	ST8	\\
20041118	&	53328.350	&	13.53	&	14.74	&	15.68	&	17.31	&	Schmidt	&	ST8	\\
20041120	&	53330.375	&	13.48	&	14.67	&	16.12	&	17.25	&	Schmidt	&	ST8	\\
20061216	&	54086.277	&	13.22	&	14.21	&	15.23	&	16.66	&	Schmidt	&	ST8	\\
20081120	&	54791.191	&	13.45	&	14.54	&	15.55	&	16.96	&	Schmidt	&	STL-11	\\
20090416	&	54938.543	&	13.26	&	14.33	&	15.44	&	16.82	&	Schmidt	&	STL-11	\\
20090519	&	54971.420	&	13.48	&	14.60	&	15.58	&	-	&	Schmidt	&	STL-11	\\
20090629	&	55011.527	&	13.28	&	14.38	&	15.41	&	16.78	&	Schmidt	&	FLI	\\
20090714	&	55027.438	&	13.60	&	14.83	&	15.92	&	17.40	&	Schmidt	&	FLI	\\
20090714	&	55027.440	&	13.58	&	14.78	&	15.91	&	17.36	&	Schmidt	&	FLI	\\
20090715	&	55028.447	&	13.56	&	14.87	&	-	&	-	&	Schmidt	&	FLI	\\
20090715	&	55028.462	&	13.56	&	14.87	&	15.94	&	-	&	Schmidt	&	FLI	\\
20090716	&	55029.453	&	13.43	&	14.64	&	15.69	&	17.07	&	Schmidt	&	FLI	\\
20090716	&	55029.469	&	13.44	&	14.63	&	15.73	&	17.10	&	Schmidt	&	FLI	\\
20090821	&	55065.354	&	13.25	&	14.38	&	15.42	&	16.85	&	Schmidt	&	FLI	\\
20090821	&	55065.377	&	13.24	&	14.36	&	15.28	&	16.84	&	Schmidt	&	FLI	\\
20090822	&	55066.284	&	13.21	&	14.25	&	15.26	&	16.59	&	Schmidt	&	FLI	\\
20091006	&	55111.424	&	13.58	&	14.79	&	15.84	&	17.20	&	Schmidt	&	FLI	\\
20091007	&	55112.379	&	13.37	&	14.57	&	15.64	&	16.92	&	Schmidt	&	FLI	\\
20091007	&	55112.390	&	13.40	&	14.60	&	15.62	&	16.91	&	Schmidt	&	FLI	\\
20091008	&	55113.354	&	13.39	&	14.61	&	15.65	&	16.92	&	Schmidt	&	FLI	\\
20091008	&	55113.377	&	13.43	&	14.63	&	15.65	&	17.10	&	Schmidt	&	FLI	\\
20091009	&	55114.246	&	13.39	&	14.54	&	15.60	&	17.01	&	Schmidt	&	FLI	\\
20091120	&	55156.238	&	13.21	&	14.30	&	15.32	&	16.67	&	Schmidt	&	FLI	\\
20091120	&	55156.259	&	13.26	&	14.37	&	15.28	&	16.66	&	Schmidt	&	FLI	\\
20091121	&	55157.267	&	13.27	&	14.36	&	15.37	&	16.70	&	Schmidt	&	FLI	\\
20091121	&	55157.287	&	13.28	&	14.40	&	15.31	&	16.65	&	Schmidt	&	FLI	\\
20100513	&	55330.355	&	13.33	&	14.48	&	15.45	&	16.91	&	Schmidt	&	FLI	\\
20100513	&	55330.374	&	13.34	&	14.42	&	15.49	&	16.93	&	Schmidt	&	FLI	\\
20100608	&	55356.418	&	13.20	&	14.29	&	15.31	&	16.65	&	Schmidt	&	FLI	\\
20100608	&	55356.438	&	13.20	&	14.30	&	15.27	&	16.68	&	Schmidt	&	FLI	\\
20100610	&	55358.468	&	13.20	&	14.26	&	15.28	&	16.68	&	Schmidt	&	FLI	\\
20100610	&	55358.488	&	13.17	&	14.28	&	15.24	&	16.65	&	Schmidt	&	FLI	\\
20100612	&	55359.508	&	13.23	&	14.32	&	15.32	&	16.74	&	Schmidt	&	FLI	\\
20100612	&	55360.442	&	13.15	&	14.23	&	15.25	&	16.59	&	Schmidt	&	FLI	\\
20100612	&	55360.456	&	13.19	&	14.31	&	15.19	&	16.64	&	Schmidt	&	FLI	\\
20100804	&	55413.309	&	13.27	&	14.34	&	15.36	&	16.57	&	Schmidt	&	FLI	\\
20100806	&	55415.397	&	13.20	&	14.30	&	15.30	&	16.65	&	Schmidt	&	FLI	\\
20100806	&	55415.416	&	13.22	&	14.34	&	15.29	&	16.67	&	Schmidt	&	FLI	\\
20100807	&	55416.360	&	13.18	&	14.26	&	15.26	&	16.66	&	Schmidt	&	FLI	\\
20100807	&	55416.380	&	13.20	&	14.30	&	15.27	&	16.67	&	Schmidt	&	FLI	\\
20100908	&	55447.520	&	13.53	&	14.74	&	15.80	&	17.31	&	Schmidt	&	FLI	\\
20100908	&	55448.413	&	13.67	&	14.88	&	15.90	&	17.40	&	Schmidt	&	FLI	\\
20100908	&	55448.435	&	13.64	&	14.85	&	15.95	&	17.36	&	Schmidt	&	FLI	\\
20100909	&	55449.480	&	13.60	&	14.84	&	15.97	&	17.48	&	Schmidt	&	FLI	\\
20101104	&	55505.273	&	13.15	&	14.22	&	15.15	&	16.54	&	Schmidt	&	FLI	\\
20101104	&	55505.298	&	13.15	&	14.20	&	15.22	&	16.55	&	Schmidt	&	FLI	\\
20101105	&	55506.288	&	13.18	&	14.31	&	15.22	&	16.64	&	Schmidt	&	FLI	\\
20101105	&	55506.313	&	13.12	&	14.25	&	15.28	&	16.70	&	Schmidt	&	FLI	\\
20110101	&	55563.256	&	13.28	&	14.44	&	15.47	&	16.92	&	Schmidt	&	FLI	\\
20110206	&	55599.228	&	13.13	&	14.26	&	15.27	&	16.58	&	Schmidt	&	FLI	\\
20110404	&	55656.398	&	13.32	&	14.41	&	15.47	&	16.70	&	Schmidt	&	FLI	\\
20110404	&	55656.422	&	13.21	&	14.26	&	15.37	&	16.79	&	Schmidt	&	FLI	\\
20110522	&	55704.392	&	13.16	&	14.21	&	15.22	&	16.57	&	Schmidt	&	FLI	\\
20110522	&	55704.411	&	13.17	&	14.25	&	15.20	&	16.54	&	Schmidt	&	FLI	\\
20110523	&	55705.335	&	13.07	&	14.13	&	15.15	&	16.47	&	Schmidt	&	FLI	\\
20110523	&	55705.354	&	13.11	&	14.19	&	15.07	&	16.52	&	Schmidt	&	FLI	\\
20110524	&	55706.324	&	13.04	&	14.12	&	15.08	&	16.33	&	Schmidt	&	FLI	\\
20110524	&	55706.343	&	13.08	&	14.17	&	15.08	&	16.23	&	Schmidt	&	FLI	\\
20110525	&	55707.335	&	13.13	&	14.23	&	15.19	&	16.53	&	Schmidt	&	FLI	\\
20110609	&	55722.354	&	13.17	&	14.25	&	15.25	&	16.57	&	Schmidt	&	FLI	\\
20110609	&	55722.371	&	13.23	&	14.31	&	15.24	&	16.41	&	Schmidt	&	FLI	\\
20110621	&	55734.470	&	13.33	&	14.52	&	15.57	&	17.02	&	Schmidt	&	FLI	\\
20110622	&	55735.475	&	13.38	&	14.54	&	15.58	&	16.97	&	Schmidt	&	FLI	\\
20110624	&	55737.404	&	13.34	&	14.47	&	15.53	&	16.92	&	Schmidt	&	FLI	\\
20110727	&	55770.412	&	13.21	&	14.35	&	15.38	&	16.72	&	Schmidt	&	FLI	\\
20110823	&	55797.366	&	13.17	&	14.29	&	15.28	&	16.64	&	Schmidt	&	FLI	\\
20110823	&	55797.386	&	13.20	&	14.31	&	15.26	&	16.63	&	Schmidt	&	FLI	\\
20110824	&	55798.363	&	13.17	&	14.29	&	15.21	&	16.63	&	Schmidt	&	FLI	\\
20110824	&	55798.390	&	13.12	&	14.24	&	15.24	&	16.53	&	Schmidt	&	FLI	\\
20110825	&	55799.370	&	13.14	&	14.28	&	15.21	&	16.58	&	Schmidt	&	FLI	\\
20110825	&	55799.392	&	13.10	&	14.22	&	15.26	&	16.60	&	Schmidt	&	FLI	\\
20110923	&	55828.302	&	13.24	&	14.39	&	15.38	&	16.83	&	Schmidt	&	FLI	\\
20111127	&	55893.206	&	13.40	&	14.59	&	15.62	&	17.12	&	Schmidt	&	FLI	\\
20111129	&	55895.296	&	13.38	&	14.58	&	15.65	&	17.01	&	Schmidt	&	FLI	\\
20111129	&	55895.316	&	13.44	&	14.63	&	15.62	&	17.09	&	Schmidt	&	FLI	\\
20111130	&	55896.277	&	13.49	&	14.71	&	15.83	&	17.27	&	Schmidt	&	FLI	\\
20111229	&	55925.222	&	13.58	&	14.82	&	15.97	&	17.41	&	Schmidt	&	FLI	\\
20111229	&	55925.236	&	13.60	&	14.85	&	15.89	&	17.41	&	Schmidt	&	FLI	\\
20120316	&	56003.488	&	13.21	&	14.37	&	15.37	&	16.74	&	Schmidt	&	FLI	\\
20120316	&	56003.509	&	13.26	&	14.40	&	15.37	&	16.80	&	Schmidt	&	FLI	\\
20120412	&	56030.484	&	13.35	&	14.49	&	15.53	&	16.94	&	Schmidt	&	FLI	\\
20120612	&	56091.443	&	13.32	&	14.46	&	15.51	&	16.95	&	Schmidt	&	FLI	\\
20120612	&	56091.462	&	13.25	&	14.34	&	15.25	&	-	&	Schmidt	&	FLI	\\
20120617	&	56096.447	&	13.25	&	14.34	&	15.25	&	-	&	Schmidt	&	FLI	\\
20120618	&	56096.522	&	13.19	&	14.30	&	15.32	&	16.64	&	Schmidt	&	FLI	\\
20120711	&	56120.420	&	13.32	&	14.28	&	15.36	&	16.64	&	Schmidt	&	FLI	\\
20120711	&	56120.447	&	13.20	&	14.27	&	15.28	&	16.60	&	Schmidt	&	FLI	\\
20120713	&	56122.421	&	13.19	&	14.25	&	15.18	&	16.55	&	Schmidt	&	FLI	\\
20120713	&	56122.443	&	13.15	&	14.23	&	15.23	&	16.47	&	Schmidt	&	FLI	\\
20120714	&	56123.438	&	13.20	&	14.29	&	15.27	&	16.58	&	Schmidt	&	FLI	\\
20120819	&	56159.393	&	13.21	&	14.34	&	15.34	&	16.71	&	Schmidt	&	FLI	\\
20120820	&	56160.374	&	13.31	&	14.44	&	15.38	&	16.84	&	Schmidt	&	FLI	\\
20120820	&	56160.398	&	13.25	&	14.38	&	15.41	&	16.80	&	Schmidt	&	FLI	\\
20120821	&	56161.453	&	13.16	&	14.25	&	15.27	&	16.54	&	Schmidt	&	FLI	\\
20120822	&	56162.378	&	13.15	&	14.24	&	15.27	&	16.63	&	Schmidt	&	FLI	\\
20120923	&	56194.364	&	13.20	&	14.26	&	15.27	&	16.61	&	Schmidt	&	FLI	\\
20121009	&	56210.265	&	13.27	&	14.39	&	15.42	&	16.76	&	Schmidt	&	FLI	\\
20121009	&	56210.284	&	13.30	&	14.42	&	15.38	&	16.74	&	Schmidt	&	FLI	\\
20121118	&	56250.355	&	13.67	&	14.91	&	16.02	&	17.41	&	Schmidt	&	FLI	\\
20121118	&	56250.369	&	13.69	&	14.92	&	15.98	&	-	&	Schmidt	&	FLI	\\
20130204	&	56328.241	&	13.70	&	14.96	&	16.11	&	17.60	&	Schmidt	&	FLI	\\
20130204	&	56328.259	&	13.72	&	15.00	&	16.03	&	17.61	&	Schmidt	&	FLI	\\
20130411	&	56394.389	&	13.48	&	14.72	&	15.84	&	17.29	&	Schmidt	&	FLI	\\
20130411	&	56394.410	&	13.54	&	14.75	&	15.82	&	17.32	&	Schmidt	&	FLI	\\
20130502	&	56415.467	&	13.31	&	14.48	&	15.46	&	16.88	&	Schmidt	&	FLI	\\
20130502	&	56415.490	&	13.30	&	14.44	&	15.49	&	16.85	&	Schmidt	&	FLI	\\
20130530	&	56443.474	&	13.32	&	14.45	&	15.44	&	16.78	&	Schmidt	&	FLI	\\
20130530	&	56443.497	&	13.25	&	14.41	&	15.42	&	16.81	&	Schmidt	&	FLI	\\
20130531	&	56444.433	&	13.32	&	14.51	&	15.46	&	16.97	&	Schmidt	&	FLI	\\
20130531	&	56444.481	&	13.30	&	14.49	&	15.53	&	16.92	&	Schmidt	&	FLI	\\
20130804	&	56509.368	&	13.50	&	14.69	&	15.79	&	17.21	&	Schmidt	&	FLI	\\
20130804	&	56509.388	&	13.55	&	14.73	&	15.73	&	17.20	&	Schmidt	&	FLI	\\
20130805	&	56510.452	&	13.65	&	14.79	&	15.89	&	17.33	&	60-cm	&	FLI	\\
20130806	&	56511.493	&	13.56	&	14.66	&	15.77	&	17.27	&	60-cm	&	FLI	\\
20130807	&	56512.481	&	13.53	&	14.62	&	15.69	&	17.16	&	60-cm	&	FLI	\\
20130808	&	56513.469	&	13.41	&	14.45	&	15.46	&	16.88	&	60-cm	&	FLI	\\
20130809	&	56514.427	&	13.41	&	14.43	&	15.45	&	-	&	60-cm	&	FLI	\\
20130904	&	56540.376	&	13.23	&	14.36	&	15.37	&	16.70	&	Schmidt	&	FLI	\\
20130904	&	56540.399	&	13.25	&	14.39	&	15.34	&	16.73	&	Schmidt	&	FLI	\\
20130905	&	56541.402	&	13.39	&	14.52	&	15.48	&	16.93	&	Schmidt	&	FLI	\\
20130906	&	56542.440	&	13.20	&	14.32	&	15.33	&	16.72	&	Schmidt	&	FLI	\\
20130911	&	56547.351	&	-	&	14.61	&	15.53	&	16.93	&	60-cm	&	FLI	\\
20130911	&	56547.471	&	13.38	&	14.41	&	15.39	&	16.92	&	60-cm	&	FLI	\\
20130914	&	56550.302	&	13.25	&	14.34	&	15.38	&	16.66	&	60-cm	&	FLI	\\
201309014	&	56550.330	&	13.24	&	14.42	&	15.40	&	16.80	&	60-cm	&	FLI	\\
20131012	&	56578.461	&	13.32	&	14.47	&	15.60	&	17.20	&	60-cm	&	FLI	\\
20131229	&	56656.327	&	14.06	&	15.46	&	16.62	&	18.10	&	Schmidt	&	FLI	\\
20140123	&	56681.265	&	13.39	&	14.52	&	15.48	&	16.84	&	Schmidt	&	FLI	\\
20140321	&	56738.483	&	13.50	&	14.75	&	15.91	&	17.19	&	Schmidt	&	FLI	\\
20140321	&	56738.504	&	13.56	&	14.78	&	15.85	&	17.29	&	Schmidt	&	FLI	\\
20140522	&	56799.515	&	13.26	&	14.41	&	15.51	&	16.78	&	Schmidt	&	FLI	\\
20140521	&	56799.542	&	13.28	&	14.48	&	15.39	&	16.83	&	Schmidt	&	FLI	\\
20140628	&	56837.443	&	13.38	&	14.49	&	15.53	&	16.83	&	Schmidt	&	FLI	\\
20140629	&	56838.421	&	13.32	&	14.46	&	15.50	&	16.82	&	Schmidt	&	FLI	\\
20140629	&	56838.478	&	13.37	&	14.50	&	15.40	&	16.81	&	Schmidt	&	FLI	\\
20140803	&	56873.307	&	13.83	&	15.14	&	16.29	&	17.89	&	Schmidt	&	FLI	\\
20140803	&	56873.328	&	13.82	&	15.14	&	16.29	&	17.71	&	Schmidt	&	FLI	\\
20140804	&	56874.317	&	13.88	&	15.18	&	16.27	&	17.83	&	Schmidt	&	FLI	\\
20140804	&	56874.351	&	13.82	&	15.13	&	16.36	&	-	&	Schmidt	&	FLI	\\
20140818	&	56888.360	&	13.91	&	15.17	&	16.22	&	17.87	&	Schmidt	&	FLI	\\
20140818	&	56888.381	&	13.90	&	15.16	&	16.34	&	-	&	Schmidt	&	FLI	\\
20140819	&	56889.295	&	13.87	&	15.13	&	16.26	&	17.83	&	Schmidt	&	FLI	\\
20140819	&	56889.318	&	13.84	&	15.15	&	16.30	&	17.75	&	Schmidt	&	FLI	\\
20140822	&	56892.358	&	13.79	&	15.02	&	16.07	&	17.55	&	Schmidt	&	FLI	\\
20140923	&	56924.325	&	13.24	&	14.35	&	15.35	&	-	&	Schmidt	&	FLI	\\
20141126	&	56988.238	&	13.37	&	14.50	&	15.72	&	16.91	&	Schmidt	&	FLI	\\
20141213	&	57005.280	&	13.37	&	14.50	&	-	&	16.80	&	Schmidt	&	FLI	\\
20141213	&	57005.303	&	13.41	&	14.57	&	15.61	&	-	&	Schmidt	&	FLI	\\
20141214	&	57006.340	&	13.38	&	14.52	&	-	&	16.84	&	Schmidt	&	FLI	\\
20141214	&	57006.363	&	13.40	&	14.56	&	15.64	&	-	&	Schmidt	&	FLI	\\
20150221	&	57074.535	&	13.50	&	14.69	&	16.03	&	17.16	&	Schmidt	&	FLI	\\
20150424	&	57136.578	&	13.66	&	14.90	&	15.83	&	17.38	&	Schmidt	&	FLI	\\
20150426	&	57138.548	&	13.67	&	14.93	&	16.07	&	17.52	&	Schmidt	&	FLI	\\
20150519	&	57162.488	&	13.42	&	14.55	&	15.85	&	16.79	&	Schmidt	&	FLI	\\
20150522	&	57164.500	&	13.39	&	14.52	&	15.83	&	16.90	&	Schmidt	&	FLI	\\
20150613	&	57186.505	&	13.79	&	15.04	&	16.50	&	17.67	&	Schmidt	&	FLI	\\
20150716	&	57220.423	&	13.89	&	15.17	&	16.53	&	17.80	&	Schmidt	&	FLI	\\
20150717	&	57221.478	&	14.01	&	15.33	&	16.75	&	17.94	&	Schmidt	&	FLI	\\
20150824	&	57259.393	&	13.85	&	15.10	&	16.29	&	17.53	&	Schmidt	&	FLI	\\
20150825	&	57260.384	&	14.10	&	15.40	&	16.61	&	17.90	&	Schmidt	&	FLI	\\
20150903	&	57269.381	&	13.73	&	14.99	&	16.25	&	17.64	&	Schmidt	&	FLI	\\
20151103	&	57330.284	&	13.32	&	14.49	&	15.68	&	16.82	&	Schmidt	&	FLI	\\
20151104	&	57331.297	&	13.33	&	14.47	&	15.74	&	16.81	&	Schmidt	&	FLI	\\
20151105	&	57332.286	&	13.34	&	14.46	&	15.76	&	16.86	&	Schmidt	&	FLI	\\
20151106	&	57333.286	&	13.39	&	14.55	&	15.79	&	16.96	&	Schmidt	&	FLI	\\
20151107	&	57334.267	&	13.38	&	14.54	&	15.78	&	16.95	&	Schmidt	&	FLI	\\
20151215	&	57372.270	&	13.29	&	14.40	&	15.66	&	16.81	&	Schmidt	&	FLI	\\
20160206	&	57425.240	&	13.36	&	14.50	&	15.41	&	16.90	&	Schmidt	&	FLI	\\
20160406	&	57485.448	&	13.40	&	14.53	&	15.47	&	16.97	&	Schmidt	&	FLI	\\
20160427	&	57506.511	&	13.53	&	14.67	&	15.72	&	17.20	&	Schmidt	&	FLI	\\
20160513	&	57522.485	&	13.17	&	14.18	&	15.14	&	16.47	&	Schmidt	&	FLI	\\
20160514	&	57523.471	&	13.04	&	14.05	&	14.99	&	16.34	&	Schmidt	&	FLI	\\
20160625	&	57565.464	&	13.17	&	14.22	&	15.19	&	16.52	&	Schmidt	&	FLI	\\
20160711	&	57581.451	&	13.20	&	14.23	&	15.22	&	16.59	&	Schmidt	&	FLI	\\
20160712	&	57582.484	&	13.22	&	14.26	&	15.25	&	16.63	&	Schmidt	&	FLI	\\
20160713	&	57583.470	&	13.21	&	14.29	&	15.24	&	16.61	&	Schmidt	&	FLI	\\
20160804	&	57605.436	&	13.19	&	14.22	&	15.16	&	16.50	&	Schmidt	&	FLI	\\
20160806	&	57607.415	&	13.18	&	14.22	&	15.17	&	16.51	&	Schmidt	&	FLI	\\
\hline \hline
\end{longtable}}

\end{document}